\documentclass[a4paper,11pt]{article}

\pdfoutput=1 

\usepackage{jcappub} 
\usepackage[T1]{fontenc} 
\usepackage[normalem]{ulem}
\usepackage{epsfig}
\usepackage{multirow}
\usepackage{color}
\usepackage{hyperref}
\usepackage{graphicx}

\def\beq{\begin{equation}} \def\eeq{\end{equation}}
\def\beqa{\begin{eqnarray}} \def\eeqa{\end{eqnarray}}

\def\AJS{{\it Ap. J. Supp.} }

\def\PL{{\it Phys. Lett.} }
\def\PR{{\it Phys. Rev.} }
\def\PRL{{\it Phys. Rev. Lett.} }

\def\RMP{{\it Rev. Mod. Phys.} }

\def\nn{\nonumber}

\title{Cosmological perturbations in theories with non-minimal coupling  between curvature and matter}

\author[a,1]{Orfeu Bertolami,\note{Also at Instituto de Plasmas e Fus\~ao Nuclear, Instituto Superior T\'ecnico.}}
\author[b]{Pedro Fraz\~ao}
\author[b]{and Jorge P\'aramos}

\affiliation[a]{Departamento de F\'{\i}sica e Astronomia, Faculdade de Ci\^encias, Universidade do Porto,\\Rua do Campo Alegre 687,
 4169-007 Porto, Portugal}
\affiliation[b]{Instituto de Plasmas e Fus\~ao Nuclear, Instituto Superior T\'ecnico, Universidade T\'ecnica de Lisboa, Av. Rovisco Pais 1, 1049-001 Lisboa, Portugal}

\emailAdd{orfeu.bertolami@fc.up.pt}
\emailAdd{pedro.frazao@ist.utl.pt}
\emailAdd{paramos@ist.edu}

\date{\today}

\abstract{In this work, we examine how the presence of a non-minimal coupling between spacetime curvature and matter affects the evolution of cosmological perturbations on a homogeneous and isotropic Universe, and hence the formation of large-scale structure. This framework places constraints on the terms which arise due to the coupling with matter and, in particular, on the modified growth of matter density perturbations. We derive approximate analytical solutions for the evolution of matter overdensities during the matter dominated era and discuss the compatibility of the obtained results with the hypothesis that the late time acceleration of the Universe is driven by a non-minimal coupling.}

\begin{document}
\maketitle
\flushbottom

\section{Introduction}

Over the past decades, the cosmological perturbation theory, resulting from the linearized Einstein field equations, has become a cornerstone of modern cosmology \cite{perturbations}. It aims to describe the large-scale structure of the Universe (LSS) \cite{SDSS} and to match the observed fluctuations of the cosmic microwave background (CMB) \cite{WMAP7}. In fact,
the Universe is, on large scales, homogeneous and isotropic since primordial times. However, small primordial density perturbations, resulting from vacuum fluctuations during inflation, grow as the Universe expands via gravitational instability ---resulting in the structures one may observe at the present time, such as galaxies and clusters of galaxies.
These matter density perturbations evolve during cosmic expansion along with metric perturbations around this smooth spacetime.

The late-time accelerated expansion of the Universe \cite{SDSS,WMAP7} can be explained either by an exotic form of energy, the so-called dark energy, or by a modification of gravity at cosmological scales. In particular, in modified theories of gravity such as $f(R)$ gravity \cite{faraoni_review}, where the linear dependence of the Einstein-Hilbert action on the Ricci
scalar is substituted by a non-linear function, it has been shown that the new terms that arise in the modified field equations can effectively lead to late-time cosmic acceleration \cite{Caretal03, CapCarTro03, NojOdi03}. However, describing the cosmological evolution of the accelerated expansion of the background is not sufficient to break down the degeneracy between modified theories of gravity and usual dark energy models --- an issue further complicated by the possibility to link $f(R)$ models with appropriate scalar-tensor theories via a conformal transformation \cite{conformal}.

The growth of matter density perturbations depends strongly on the underlying theory of gravity (namely General Relativity (GR) or a modification of it), since the modified field equations lead to different evolutions of cosmological perturbations and alter the characteristic imprints that these leave in the CMB and in the matter power spectrum inferred from galaxy clusters. It is therefore an outstanding probe to modified gravity models, allowing for the discrimination between the latter and usual
dark energy models (which assume the validity of GR) \cite{White, Shirata:2007qk, Koivisto:2006ie, Tsujikawa:2007gd,Tsujikawa_b,Sealfon}. For instance, in the context of $f(R)$ theories, it is not possible to describe simultaneously small scales of galaxy surveys data and large scales of CMB data, as the latter are suppressed in comparison with the former \cite{Bean}. The cosmic microwave background high-precision data can also be used to constrain $f(R)$ gravity models \cite{Amendola,Hu, Li1, Li2, Wei, Starobinsky, Carloni:2007yv}.

By the same token, the evolution of matter perturbations can also lead to stringent restrictions on $f(R)$ theories that reproduce the Einstein-Hilbert action at high curvatures through a completely general procedure \cite{Song, Carroll, Dombriz}: in particular, the linear cosmological evolution of $f(R)$ gravity can reduce the large-angle CMB anisotropy, as well as induce a different profile for the linear matter power spectrum and the correlations between the Integrated-Sachs-Wolfe effect and the angular power spectra of galaxies \cite{Song}.

Another interesting gravity probe is gravitational lensing \cite{MTG-lensing}: indeed, the relation between the gravitational
potentials in the metric, which are responsible for gravitational lensing, and the matter overdensities, depends on the theory of gravity \cite{Zhang2}.

In the context of the non-minimal coupling between scalar curvature and matter \cite{coupling}, it has been shown in several works that both dark energy and dark matter can be effectively described within this framework, along with other cosmological and astrophysical implications. For instance, the mimicking of dark matter \cite{mimic,mimic_a} or even dark energy \cite{expansion}, the reheating scenario after inflation \cite{coupling_inflation}, and the change in the gravitational potential \cite{potential}.

In this work, one examines the possibility of a non-minimal coupling between curvature and matter via the study of the evolution of cosmological perturbations, by following the implications of the modified field equations and the non-conservation of the energy-momentum tensor.

This work is organized as follows: first, one introduces the non-minimal coupling model. Then one shortly reviews the
formalism of the cosmological perturbation theory and obtains the linearized equations in the presence of a non-minimal coupling. In the last sections, one derives the equations of motion for scalar perturbations and matter overdensities in the subhorizon approximation and studies the respective evolution for the relevant modes in the formation of LSS along with the respective conclusions. Finally, conclusions are drawn as to the main effects stemming from the presence of a non-minimal coupling.

\section{The non-minimal coupling}

\subsection{Modified field equations}

A recent generalization of modified theories of gravity resorts to a non-minimal curvature-matter coupling, further extending the presence of non-linear functions of the scalar curvature in the action functional \cite{coupling}. The considered action is

\begin{equation}
S=\int d^{4}x\sqrt{-g}\left[ {1 \over 2}f_{1}(R)+\left[1+f_{2}(R)\right]\mathcal{L}_{m}\right]~~,
\label{model}
\end{equation}
where $f_{i}(R)$ (with $i=1,2$) are arbitrary functions of the scalar curvature and $g$ is the metric determinant. Setting $f_{2}(R)=0$, one obtains the usual $f(R)$ theories of gravity, and GR is recovered if in addition the linear function $f_{1}(R)= R/\kappa$ is considered (with $\kappa = 8\pi G_{}$), where $G_{}$ is the Newtonian gravitational constant.

The field equations are obtained from the variation of Eq. (\ref{model}) with respect to the metric \cite{coupling},

\begin{eqnarray}
&&FR_{\nu}^{\mu}- {1 \over 2}\delta_{\nu}^{\mu}f_{1}-\left(g^{\mu\rho}\nabla_{\rho}\nabla_{\nu}-\delta_{\nu}^{\mu}\square\right)F=\left(1+f_{2}\right)T_{\nu}^{\mu}~~,\label{field_equations}
\end{eqnarray}

\noindent where $T_{\nu}^{\mu}$ is the energy-momentum tensor, $F_i \equiv df_i/dR$ and $F\equiv F_{1}-2F_{2}\rho$, with $\rho$ the energy density. One assumes a matter Lagrangian of the form ${\cal L}_{m}=-\rho$ (see Refs. \cite{fluid} for a thorough discussion).

Taking the trace of Eq. (\ref{field_equations}), one gets

\begin{eqnarray}
&&FR-2f_{1}+3\square F = \left(1+f_{2}\right)T~~,\label{trace}
\end{eqnarray}

\noindent where $T$ is the trace of the energy-momentum tensor. In the following, one considers that the matter-energy content of the Universe is well described by a perfect fluid,  $T^{\mu\nu}=(\rho+p)u^{\mu}u^{\nu}+pg^{\mu\nu}$, 
where $u^{\mu}$ is the four-velocity field in comoving coordinates (with $u^{0}=1$ from the condition $u^{\mu}u_{\mu}=-1$).

For an homogeneous and isotropic spacetime, described by a Friedmann-Robertson-Walker (FRW) metric, the background evolution of the field equations for the non-minimal coupling, Eq. (\ref{field_equations}), takes the form 

\begin{eqnarray}
3FH^{2}&=&\rho+\rho_c~~,\label{modified_friedmann}\\
-F (2\dot{H}+3H^{2} )&=&p+p_c~~,
\end{eqnarray}

\noindent where the new terms are due to the modification of gravity and appear as curvature components of the energy-density and pressure, which are defined as

\begin{eqnarray}
\rho_c&\equiv&-{1\over 2}\left(f-FR\right)-3H\dot{F}~~,\\
p_c&\equiv&f_{2}\,p+ (\ddot{F}+2H\dot{F} )+{1\over 2}\left(f_{1}-FR\right)~~,
\end{eqnarray}

\noindent where the dots refer to time derivatives and $f\equiv f_1-2f_2\rho$.

\subsection{Non-conservation of the energy-momentum tensor}

One distinct signature of the non-minimal coupling is the non-conservation of the energy-momentum tensor \cite{coupling}. Using the Bianchi identities, $\nabla_{\mu}G_{\nu}^{\mu}=0$, and taking the covariant derivative of the field equations, Eqs. (\ref{field_equations}), results in

\begin{equation}
  \nabla_{\mu}T_{\nu}^{\mu}=- \left(\rho\delta_{\nu}^{\mu}+T_{\nu}^{\mu}\right)\nabla_{\mu}\log\left(1+f_2\right)~~,\label{non-cons-2}
\end{equation}
which one may identify as an energy-momentum exchange between matter
and spacetime geometry. Setting $f_2(R)=0$ leads to the usual GR covariant conservation of the energy-momentum
tensor.

However, in a perturbed spacetime, these covariant derivatives will differ from that of the background metric due to the perturbation on the Christoffel symbols, and one must also take into account the perturbations on the \emph{r.h.s.} of Eq. (\ref{non-cons-2}), which are non-vanishing due to the presence of the non-minimal coupling.

In this context, for a perfect fluid description, the above non-conservation of the energy-momentum tensor, Eq. ($\ref{non-cons-2}$), leads to the extra force \cite{coupling}

\begin{eqnarray}
f^{\mu} & = &  {1 \over \rho+p}\left[\left(-\rho+p\right)\nabla_{\nu}\log\left(1+f_2\right)+\nabla_{\nu}p\right]h^{\mu\nu}~~,
\label{extra-force}
\end{eqnarray}

\noindent where $h^{\mu\nu}=g^{\mu\nu}+u^\mu u^\nu$ is the projection operator. This is orthogonal to the fluid four-velocity, since $f^{\mu}u_\mu\sim h^{\mu\nu}u_\mu=0$.

\section{Cosmological perturbations}

The theory of cosmological perturbations resorts to the evolution of small fluctuations around an idealized homogeneous and isotropic background Universe. These cosmological perturbations can be decomposed in three types: scalar, vector and tensorial modes, which have different and independent evolutions during cosmic expansion. In general, only scalar perturbations present instabilities that grow during the expansion of the Universe and lead to the formation of LSS. Conversely, vector and tensorial modes do not present such growing inhomogeneities: the former decay with the
cosmological expansion and the latter do not present any coupling with the energy density or pressure because the non-diagonal terms of the energy-momentum tensor are zero. Hence, one restricts the attention to the scalar mode of cosmological perturbations.

The linearized modified field equations and the non-conservation of the energy-momentum tensor constitute the set of equations which govern the evolution of cosmological perturbations around the smooth FRW background. One now considers the most general form of the line element for the perturbations of the FRW spacetime, which in the longitudinal gauge yields the line element \cite{perturbations},

\begin{eqnarray}
ds^{2} & = & -(1+2\Phi)\, dt^{2}+a^{2}(t)(1-2\Psi)\delta^i_jdx_{i}\, dx^{j}~~,
\label{metric}
\end{eqnarray}

\noindent where the perturbations coincide with the Bardeen gauge invariant potentials.

The linearized field equations, Eq. (\ref{field_equations}), and the non-conservation of the energy-momentum tensor, Eq. (\ref{non-cons-2}) lead, to first order, to a set of equations governing the cosmological dynamics of the potentials $\Phi$ and $\Psi$ along with the matter density and pressure perturbations, as one will derive in the following paragraph. For a further discussion on a more general coupling function of the matter Lagrangian see Ref. \cite{nesseris}.

\subsection{Linearized field equations}

The linearization of the field equations is obtained considering the first order perturbation in the curvature due to metric fluctuations.  Therefore, the time-time component of the field equations, Eq. (\ref{field_equations}), is
 
\begin{align}
-\left(1+f_{2}\right)&\,\delta\rho-\left(F_{2}\rho\right)\,\delta R=\nn\\
=&\left(-2F_{2}\rho\right)\left[-3\ddot{\Psi}-3H(\dot{\Phi}+2\dot{\Psi})-3 (\dot{H}+H^{2} )\left(2\Phi\right)-{1\over a^{2}}\nabla^2\Phi+3 (\dot{H}+H^{2} )\,\delta\ln\left(F_{2}\rho\right)\right]\nn\\
&+F_{1}\left[6H (\dot{\Psi}+H\Phi)-{1\over a^{2}}\nabla^2\left(2\Psi\right)+3 (\dot{H}+H^{2} )\,\delta\ln F_{1}\right]\nn\\
&+\left[6H\dot{F}\Phi+ \dot{F}(\dot{\Phi}+3\dot{\Psi} )-3H\;\dot{\delta F}+{1\over a^{2}}\nabla^2\left(\delta F\right)\right]~~,
\label{mfe_00}
\end{align}

\noindent while the spatial-spatial component reads

\begin{align}
\delta_{j}^{i}&\left[\left(1+f_{2}\right)\delta p\,+p\,\delta f_{2}\right]=\nn\\
&=\delta_{j}^{i}\left(-2F_2\rho\right)\left[-\ddot{\Psi}- (\dot{H}+3H^{2} )\left(2\Phi\right)-H (\dot{\Phi}+6\dot{\Psi} )+{1\over a^{2}}\nabla^2\Psi\right]\nn\\
&+\delta_{j}^{i}\left[\square F+ (\dot{H}+3H^{2} )\left(\delta F\right)\right]-{F\over a^{2}}\delta^{ik}\nabla_{k}\left[\nabla_{j}\left(\Phi-\Psi+\delta \ln F\right)\right]~~.
\label{mfe_ij}
\end{align}

\noindent Here, one identifies three types of terms in the \emph{r.h.s.} of these equations: one due to the presence of the coupling function, $f_2(R)$, other stemming from the $f_1(R)$ non-linear curvature term, and a third involving a linear combination of these, present in the function, $F=F_1-2F_2\rho$, and respective fluctuation $\delta F$.

\subsection{Matter density perturbations}

After considering the linearization of the modified field equations, one now examines the linear approximation of the components of the non-conservation of the energy-momentum tensor, Eq. (\ref{non-cons-2}). First, it is relevant to introduce a new function which may be identified in the extra force given in Eq. (\ref{extra-force}) for a pressureless perfect fluid 

\begin{eqnarray}
f^{\mu} & = &  -\left(\partial_\nu\Phi_c\right)h^{\mu\nu}~~,
\end{eqnarray}

\noindent where one defines a coupling potential 

\begin{eqnarray}
\Phi_c\equiv \log\left(1+f_2\right)~~, 
\label{coupling_potential}
\end{eqnarray}
 
\noindent {\it i.e.} the extra force $f^\mu$ is the gradient of this potential. Scalar fluctuations of the metric will lead to a corresponding perturbation in this potential, $\delta \Phi_c=\log^\prime\left(1+f_2\right)\,\delta R$.

The equation of motion for matter overdensities results from the linearization of the non-conservation of the energy-momentum tensor, Eq. (\ref{non-cons-2}): the \emph{r.h.s.} of the time component is zero, while the spatial components read

\begin{eqnarray}
\nabla_{\mu}T_{\phantom{\;}i}^{\mu}&=&-\left(\delta_{i}^{\mu}\rho_m+T_{i}^{\mu}\right)\partial_{\mu}\Phi_{c}~~, 
\label{deltaTij}
\end{eqnarray}

\noindent where one takes the case of a Universe dominated by matter, $\rho\approx \rho_m$. Introducing the velocity potential $v$, defined through $u_{i} \equiv-\partial_{i}v$, one may calculate from the above expression the first order perturbation in the components $\delta T^i_0$ of the energy-momentum tensor

\begin{eqnarray}
\dot{v}-\dot{\Phi}_{c}v&=&\Phi-\delta\Phi_{c}~~.
\label{matter_first}
\end{eqnarray}

Thus, the time and spatial components of the non-conservation of the energy-momentum tensor, Eq. (\ref{non-cons-2}) lead to a set of coupled differential equations which determine the evolution of the matter fluctuations, $\delta\rho_m$:

\begin{eqnarray}
{\delta\dot{\rho}_{m}\over \rho_{m}}+3H {\delta\rho_{m}\over \rho_{m}}&=&3\dot{\Psi} -{k^{2}\over a^{2}}v_{m}~~,\label{emt1}\\
\dot{\upsilon}_{m}+(H+\dot{\Phi}_{c})\upsilon_{m}&=&{1\over a}(\Phi+\delta\Phi_{c})~~\label{emt2},
\end{eqnarray}

\noindent where the last equation is obtained by introducing the velocity potential, $v_{m}=v/a$, in Eq. (\ref{matter_first}).
These equations, along with the perturbed field equations, Eqs. (\ref{mfe_00}) and (\ref{mfe_ij}), constitute the set of equations that govern the cosmological evolution of scalar perturbations in the longitudinal gauge.

\subsection{Modification of gravity}

Having derived the first order perturbations of the field equations along with the matter density perturbations, it is now possible to study how the relevant quantities in the subhorizon approximation deviate from the usual $f(R)$ theories and GR. Firstly, one shows that the identity $\Phi=\Psi$ no longer holds: since the non-diagonal components of the energy-momentum tensor vanish, $\delta T^i_{\;j}=0$, the last term on the \emph{r.h.s.} of the spatial-spatial component of the modified field equations, Eq. (\ref{mfe_ij}), must vanish for $i\neq j$ \cite{perturbations}. This leads to a condition involving the potentials and which, for the present model, has an additional term due to the modification of gravity. Recalling that $\delta$ denotes terms which are linear both in the metric and matter fluctuations such that

\begin{eqnarray}
\delta F&=&\delta\left(F_{1}-2F_{2}\rho\right)=\left({dF_{1}\over dR}-2{dF_{2}\over dR}\rho\right)\delta R-2F_{2}\,\delta\rho~~, 
\label{deltaF}
\end{eqnarray}

\noindent one gets

\begin{eqnarray}
\Phi-\Psi=-\delta \ln F=-\delta\ln\left(F_{1}-2F_{2}\rho\right)~~,
\label{potentials}
\end{eqnarray}

\noindent which reduces to the usual GR result $\Phi=\Psi$ in the absence of a modification of gravity.   

For the formation of LSS, the relevant scales correspond to modes deep inside the horizon during the matter dominated era, since they reenter the horizon before matter-radiation equality is attained. These obey the condition $k^{2}/a^{2}H^{2}\gg 1$ and, noting that during the matter-dominated era one has $R\sim H^{2}$, one concludes that ${k^{2}/a^{2}R}\gg1$. Within this regime, one may neglect time derivatives of the potentials and retain only terms on the energy density fluctuations \cite{Tsujikawa:2007gd,Bean} (further discussion on the validity of the subhorizon approximation is found in Refs. \cite{Dombriz,Diego}).

In this approximation, the fluctuation of the scalar curvature in Fourier space becomes

\begin{eqnarray}
\delta R\approx-2{k^{2}\over a^{2}}\left(2\Psi-\Phi\right)\;\;.
\label{deltaR} 
\end{eqnarray}

\noindent For GR, the time-time component of the field equations, $(k/ a)^{2}(2\Psi/\kappa)\approx -\delta\rho$, along with the relation $\Phi=\Psi$ for the potentials leads to the standard generalized Poisson equation of the form, $\Phi\approx-4\pi G_{}(a/k)^2 \delta \rho$, where $\Phi$ is regarded as a generalized Newtonian potential. For a modified theory of gravity, one has a new coupling constant $\tilde{G}$, which differs from $G_{}$ due to the presence of factors  involving the functions $f_i(R)$, as shown below.
Indeed, the equation for the potential $\Phi$ in the subhorizon approximation has the form of a Poisson equation, 

\begin{eqnarray}
\Phi\approx-4\pi \tilde{G}{a^2\over k^2}\delta\rho~~, 
\label{Poisson_phi}
\end{eqnarray}

\noindent where one identifies a modified gravitational coupling constant, $\tilde{G}$, which differs from $G$ due to the modification of gravity.

The expression for $\tilde{G}$ is obtained from the time-time component of the modified field, Eq. (\ref{mfe_00}), which, in this approximation, becomes

\begin{eqnarray}
-\left(1+f_{2}\right)\,\delta\rho-\left(F_{2}\rho\right)\,\delta R&=&-{k^{2}\over a^{2}} [-F_{1}\left(2\Psi\right)+\left(2F_{2}\rho\right)\Phi+\left(\delta F\right) ]~~,
\label{S00a}
\end{eqnarray}

\noindent where $\delta R$ is given by Eq. (\ref{deltaR}). The fluctuation, $\delta F$, in Eq. (\ref{deltaF}), can be written as a function of the potential, $\Phi$, and the matter density fluctuation, $\delta \rho$, from the perturbation on the scalar curvature, Eq. (\ref{deltaR}). Using this result in the time-time component of the field Eqs. (\ref{mfe_00}), with the subhorizon approximation, Eq. (\ref{S00a}), leads to an expression for the modified gravitational coupling constant 

\begin{eqnarray}
 {\tilde{G}\over G_{}}\equiv {1+2 {k^{2}\over a^{2}R}\left(2m_{1}-m_{2}\right)\over 1+3 {k^{2}\over a^{2}R}m_{1}}\Sigma~~,
\label{Gtilde}
\end{eqnarray}

\noindent where one identifies the quantities

\begin{eqnarray}
\Sigma &\equiv & {1+f_{2}\over \kappa F}={1+f_{2}\over \kappa (F_1-2F_2\rho)}~~,
\label{sigma}
\end{eqnarray} 

\noindent and

\begin{eqnarray}
m_{1}&\equiv&R{F^\prime \over F}=R{F^\prime_{1}-2F^\prime_{2}\rho \over F_{1}-2F_{2}\rho}~~,\label{m1}\\
m_{2}&\equiv&R{F_{2}\over 1+f_2} ~~.\label{m2}
\end{eqnarray}

\noindent As expected, inserting $f_1(R)= R/\kappa$ and $f_2(R) = 0$ yields $m_1=m_2 = 0$ and one recovers the usual gravitational constant, $\tilde{G}=G_{}$. Notice that, for $f(R)$ theories, $m_2$ vanishes: this parameter thus emerges as a new result due to the effect of the non-minimal coupling.

From the relation between the potentials, Eq. (\ref{potentials}), the equation for the second potential has also a Poisson form,

\begin{equation}
\Psi\approx -4\pi G_{} q{a^{2} \over k^{2}}\delta\rho~~,
\label{Poisson_psi}
\end{equation}

\noindent and the deviation from GR is now reflected in the quantity 

\begin{eqnarray}
q&=&{1+2{k^{2}\over a^{2}R}\left(m_{1}+m_{2}\right)\over 1+3{k^{2}\over a^{2}R}m_{1}}\Sigma~~.
\label{q}
\end{eqnarray}

As already mentioned, the impact on weak lensing provides a straightforward test to these modified theories of gravity. The potential that determines the weak lensing is given by $\Phi_{WL}\equiv \Phi+\Psi$. Using the Poisson equations for the potentials, one may rewrite Eq. (\ref{sigma}) in the suggestive form $\Sigma=(\tilde{G}/G_{}+q)/2$, so that that one may write $\Phi_{WL}\simeq -8\pi G_{} (a/k)^2\delta\rho\,\Sigma$ \cite{MTG-lensing}, showing the connection between $\Sigma$ and a potential deviation from GR weak lensing.

By the same token, it is also possible to construct the anisotropic parameter $\eta\equiv (\Phi-\Psi)/\Psi$, which also indicates a deviation from GR ($\eta=0$),

\begin{eqnarray}
\eta&=& {2{k^{2} \over a^{2}R}\left(m_{1}-2m_{2}\right)\over 1+2 {k^{2}\over a^{2}R}\left(m_{1}+m_{2}\right)} ~~.
\label{eta}
\end{eqnarray}

\subsection{Comparison with $f(R)$ theories of gravity}

After obtaining the main results induced by the modification of gravity, it is now possible to make a comparison between the usual $f_1(R)$ theories and the non-minimal coupling with an additional $f_2(R)$ coupling function. 

First, in the equations for the modified field equations, Eqs. (\ref{mfe_00}) and (\ref{mfe_ij}), it is clear that the presence of the non-minimal coupling results in a modification of the \emph{l.h.s.}, which is a characteristic of this model, and two modifications in the \emph{r.h.s.} of the same equations, which consist in the inclusion of a new term dependent only on the coupling function and an extension of the term on the $F_1(R)$ function to the linear combination, $F=F_1-2F_2\rho$. This last extension appears throughout the obtained results and, in particular, in the difference of the potentials in Eq. (\ref{potentials}).

The other relevant distinct property resorts to the gravitational interaction depicted by the $\Sigma$ quantity, Eq. (\ref{sigma}), and the parameters $m_1$ and $m_2$, given by Eqs. (\ref{m1}) and (\ref{m2}). These determine the form of the gravitational interaction, $\tilde{G}$, the factor, $q$, on the modified Poisson equation for the potential, $\Psi$, in Eqs. (\ref{Gtilde}) and (\ref{q}), respectively, and the anisotropic parameter, $\eta$, in Eq. (\ref{eta}). In comparison with $f_1(R)$ theories, the coupling function results in a modification of $m_1$ and $\Sigma$, and the inclusion of a competing term $m_2$.

For the $\Sigma$ quantity it is clear the same extension of the $F_1$ function to the combination $F=F_1-2F_2\rho$, and the presence of the $f_2(R)$ coupling function, such that, if one has a dominant coupling function, $f_2(R)\gg 1$, then

\begin{eqnarray}
\Sigma & \approx & {f_{2}\over \kappa (F_1-2F_2\rho)}~~,
\end{eqnarray} 

\noindent and one concludes that the presence of the coupling function overcomes the usual $f_1(R)$ theories of gravity in the weak lensing behaviour.

Then, for the parameter $m_1$, Eq. (\ref{m1}), one has also the same extension for the function $F_1$, but the parameter $m_2$ corresponds to an emergent result in this new theory of gravity, and indicates the dominance of the coupling function. Hence, it is clear that the parameters $m_i$ are responsible for different regimes of dominance of the $f_i(R)$ functions, and it is $\Sigma$ which in fact induces the deviation from GR in both theories. 

At last, in the context of the evolution of matter overdensities, the usual $f_1(R)$ theories of gravity only induce the already seen modification in the gravitational interaction, $\tilde{G}$. However, the inclusion of the $f_2(R)$ coupling, besides the modification of gravity as already seen, it also induces a new behaviour from the presence of the extra-force, which was not present in the former. This effect leads to the presence of the coupling potential, Eq. (\ref{coupling_potential}), and respective fluctuation, $\delta\Phi_c$, in the equation for matter density perturbations, Eq. (\ref{emt2}). In the following section one will see the physical implications of such terms in the cosmological dynamics of matter overdensities.

\section{Evolution of matter overdensities}

\subsection{Equation of motion in the subhorizon approximation}

For the formation of LSS one must consider the evolution of matter density perturbations from the set of equations, Eqs. (\ref{emt1}) and (\ref{emt2}). If one introduces the gauge-invariant matter density perturbation, $\delta_m\equiv \delta \rho_m/\rho_m +3H v$, the respective equation for the evolution of matter overdensities, in Fourier space, becomes

\begin{eqnarray}
\ddot{\delta}_{m}+(2H+\dot{\Phi}_{c})\dot{\delta}_{m}+{k^{2}\over a^{2}}\left(\Phi+\delta\Phi_{c}\right)&=&3 \left[\ddot{B}+(2H+\dot{\Phi}_{c})\dot{B}\right]~~,
\label{delta_complete}
\end{eqnarray}
\noindent where $B\equiv \Psi+Hv$.
This result shows that the presence of the non-minimal coupling induces the background term, $\dot{\Phi}_c$, and a sort of backreaction effective perturbative term, $\delta \Phi_c$, due to the presence of the extra force. In this case, the subhorizon approximation consists in neglecting the terms on the time derivatives of $B$, so that the \emph{r.h.s.} of the above expression may be disregarded.

The term on $\dot{\delta}_{m}$, which is due to the expansion of the Universe in the absence of the coupling, has a frictional effect: its absence would lead to an exponential growth of matter density perturbations. In the presence of the coupling function there is a new similar term which shows up in the equation, $\dot{\Phi}_c=f_2\dot{R}/ [(1+f_2)R]$. This indicates that the non-minimal coupling acts to reinforce (or decrease) the effect of the background expansion of the Universe.

To obtain further insight about this term, it is relevant to consider the space component of the geodesic equation for a particle in the presence of the extra force,

\begin{equation}\frac{du^{\alpha}}{ds}+\Gamma_{\mu\nu}^{\alpha}u^{\mu}u^{\nu}=f^{\alpha}~~.\end{equation}

\noindent At zeroth order, this component reads

\begin{eqnarray}
{du^{i}\over ds}&=&-(2H+\dot{\Phi}_{c})u^{0}u^{i}~~,
\label{frictional}
\end{eqnarray}

\noindent clearly showing the similarity between the frictional behaviours of the coupling function and the one from the expansion of the Universe. This behaviour accounts for the adiabatic expansion of the Universe without an interchange of energy between matter and curvature, in particular in what concerns the evolution of matter overdensities.

To understand the origin of the second other term, $\delta \Phi_c$, in Eq. (\ref{delta_complete}), one must consider the subhorizon approximation. Since $\delta \Phi_c \sim \delta R$, and from the perturbation on the scalar curvature, $\delta R$, given by Eq. (\ref{deltaR}), one may also identify a Poisson equation as in the case of the potentials $\Phi$ and $\Psi$. Using the equations for these potentials, Eqs. (\ref{Poisson_phi}) and (\ref{Poisson_psi}), and the modified coupling constant, $\tilde{G}$, in Eq. (\ref{Gtilde}), one obtains

\begin{eqnarray}
\delta \Phi_c=-4\pi G_c {a^2\over k^2}\delta\rho_m~~, 
\label{delta_phi_c}
\end{eqnarray}

\noindent where the coupling constant, $G_c$, is given by

\begin{eqnarray}
{G_c\over G_{}}&\equiv& {-{k^{2}\over a^{2}R}m_{2}\left(1+6{k^{2}\over a^{2}R}m_{2}\right)\over 1+3 {k^{2}\over a^{2}R}m_{1}}\Sigma~~.
\label{Gc}
\end{eqnarray}

\noindent Therefore, Eq. (\ref{delta_phi_c}) reveals that the extra force, through the perturbation in the coupling potential, acts as an effective gravitational interaction for the growth of matter density perturbations. Here one observes the dependence on the quantity $m_2$: in the absence of the coupling function $m_2=0$ and the extra force vanishes.

With this identification, the equation of motion for matter overdensities, Eq. (\ref{delta_complete}), becomes 

\begin{eqnarray}
\ddot{\delta}_{m}+(2H+\dot{\Phi}_c)\dot{\delta}_{m}-4\pi (\tilde{G}+G_c)\rho_m\delta_{m}\simeq 0 ~~.
\label{delta_m}
\end{eqnarray}

From the above, one sees that the non-minimal coupling also induces a sort of backreaction gravitational effect on the evolution of matter overdensities from the action of the extra force.
From the identification with an effective gravitational constant, $G_c$, the perturbation on the coupling potential, $\delta \Phi_c$, due to the presence of matter density perturbations, $\delta\rho_m$, also induces an attractive interaction in the matter perturbations in a way which increases the growth of matter overdensities similar to the one of Newtonian gravity.

Using the explicit expressions for the coupling constants, Eqs. (\ref{Gtilde}) and (\ref{Gc}),
one has a total gravitational interaction, $G_{eff}\equiv \tilde{G}+G_c$, given by

\begin{eqnarray}
{G_{eff}\over G_{}}&=& {1+4{k^{2}\over a^{2}R}\left[m_{1}-m_{2}\left(1+3{k^{2}\over a^{2}R}m_{2}\right)\right] \over 1+3{k^{2}\over a^{2}R}m_{1}}\Sigma~~.
\end{eqnarray}

\noindent Through this expression, one may conclude that it is the quantity $\Sigma$ that in fact characterizes the modification of gravity, which is independent of the mode under consideration, and the expressions for $m_i$ indicate whether this is due to a modification of gravity, $\tilde{G}$, or rather the induced extra force, $G_c$. Although the above splitting of the effect of the non-minimal coupling into a direct gravitational impact plus the action of an extra force can be regarded as somewhat arbitrary, it is very useful to better grasp the phenomenological implications of the model under scrutiny, and to physically interpret the behaviour of competing terms.

\subsection{Dynamics for a power-law coupling function}

To understand the dynamics of matter perturbations in the presence of the non-minimal coupling, one must study the growth behaviour of matter overdensities from the respective equation of motion, Eq. (\ref{delta_m}), where the departure from the standard behaviour comes from the presence of the $\dot{\Phi}_c$ frictional term and 
the modification of gravity with the effective attractive interaction from the extra force, $G_c$, resulting in the total gravitational coupling constant $G_{eff}$.  

Following previous works, one takes the specific coupling function of a power-law form, $f_2(R)=(R/R_2)^n$, where $R_2$ is a characteristic curvature scale. For this term to be relevant only at late times, and thus drive the accelerated expansion of the Universe, the exponent $n$ has to be negative and $R_2$ smaller than the current value of the scalar curvature \cite{expansion}. Depending on the value of $R_2$, the additional terms derived from the non-minimal coupling may be dominant or perturbative at the time scale relevant for structure formation. These two options are considered in the following discussion.

For this class of functions, the $\Sigma$-quantity, Eq. (\ref{sigma}), becomes

\begin{eqnarray}
\Sigma &=& {1+f_{2}\over 1-2n{\kappa \rho_{m}\over R} f_{2}}~~.
\label{sigma_approx}
\end{eqnarray}

\noindent Considering a perturbative coupling function, $f_2\ll 1$, along with a perturbative behaviour in a way that the derivative terms in the trace Eq. (\ref{trace}) do not have a significant contribution, one may consider $R\sim \kappa \rho$ and the previous expression, Eq. (\ref{sigma_approx}), can be approximated by

\begin{eqnarray}
\Sigma&\approx&{1+f_{2} \over 1-2nf_{2}}\approx1+\left(1+2n\right)f_{2}\sim 1~~,
\label{sigma1}
\end{eqnarray}

\noindent and, for a power-law coupling function with a negative exponent along with the condition, $n<-1/2$, one has $\Sigma<1$. For a dominant coupling function one has instead 

\begin{eqnarray}
\Sigma\approx -{1\over 2n}{{R\over \kappa\rho} }~~.
\label{sigma2}
\end{eqnarray}

\noindent This expression, along with the restriction $\Sigma>0$ for a correct cosmological behaviour, places a strong condition for negative values for the exponent, $n<0$, if the coupling dominates.

For the quantities $m_1$ and $m_2$, one has

\begin{eqnarray}
m_{1}&=&{2n\left(1-n\right)\frac{\kappa\rho_m}{R}\; f_{2}\over 1-2n\frac{\kappa\rho_m}{R}f_{2}}\approx
\begin{cases}
{2n\left(1-n\right)\frac{\kappa\rho_m}{R}\; f_{2}} \;,\;\; f_{2}\ll1 \cr \cr
{n-1}  \;,\;\; f_{2}\gg1
\end{cases}
~~,
\label{m1_dominant}
\end{eqnarray}

\noindent and

\begin{eqnarray}
m_{2}&=&n{f_{2}\over 1+f_{2}}\approx
\begin{cases}
n{f_{2}}  \;,\;\; f_{2}\ll1 \cr \cr
 n  \;,\;\; f_{2}\gg1
\end{cases}~~,
\end{eqnarray}

\noindent where the corresponding values, in general negative, increase from small values when the coupling function is perturbative, to become of order unity in the non-perturbative regime. 

For the present case, the corresponding potential can be approximated as

\begin{eqnarray}
\dot{\Phi}_{c} & =& n{f_{2}\over 1+f_{2}}{\dot{R}\over R}\approx
\begin{cases}
nf_{2}\frac{\dot{R}}{R}  \;,\;\; f_{2}\ll1 \cr \cr
 n\frac{\dot{R}}{R}  \;,\;\; f_{2}\gg1
\end{cases}~~,
\label{phi_c}
\end{eqnarray}

\noindent using $F_2=nf_2/R$ for a power-law function. Since the scalar curvature decreases during
cosmic expansion, one concludes that this term acts always as a frictional term, $\dot{\Phi}_{c}>0$, for a negative exponent.

\subsection{Cosmic evolution}

In general, the coupling term, responsible for the cosmic expansion, should arise only at late-times, such that, at the beginning of the matter dominated era, one would have a small deviation of GR with $m_i\ll 1$ and $m_i(k^2/a^2 R)\,\ll 1$. As cosmic expansion evolves, not only does the first factor, ${k^2/ a^2 R}$, becomes approximately constant, but so do the parameters $m_i$. However, the factor $k^2/a^2 R$ becomes very large and $m_i$ of the order unity: therefore, one should consider two different regimes for the dominance of the coupling function along with the factors $m_i(k^2/a^2 R)$. For a deviation of GR only due to the non-minimal coupling one considers $f_1(R)=R/\kappa$ and a power-law coupling function $f_2(R)\sim R^n$.

The equation of motion, Eq. (\ref{delta_m}), can be rewritten in terms of the scale factor

\begin{eqnarray}
\delta_{m}^{\prime\prime}+{3\over 2a}\left(1-\omega_{eff}\right)\delta_{m}^{\prime}-{12\pi G_{eff}\over a^{2}}\Omega_{m}F\,\delta_{m}&=&0~~,
\label{delta_a}
\end{eqnarray}

\noindent where a prime stands for derivative with respect the scale factor and one defines 

\begin{eqnarray}
\Omega_m\equiv {\rho_m\over 3H^{2}\left(F_{1}-2F_{2}\kappa\rho_{m}\right)}~~,
\end{eqnarray}

\noindent and an effective parameter 

\begin{eqnarray}
\omega_{eff} \equiv -1-{2\over 3}a\left({H^{\prime}\over H}+\Phi_{c}^{\prime}\right)~~. 
\end{eqnarray}

\subsubsection{Matter dominated era}

At the beginning of the dominated matter era, for a given perturbative coupling function, one should have ${k^2/ a^2 R}\;m_i\ll 1$, so that the total gravitational constant can be approximated as

\begin{equation}
{G_{eff}\over G_{}} \approx \left[1+{k^{2}\over a^{2}R}\left(m_{1}-4m_{2}\right)\right]\Sigma \approx\left[1-2n\left(1+n\right){k^{2}\over a^{2}R}f_{2}\right]\Sigma~~,
\end{equation}

\noindent since the parameters $m_i$ are given by $m_{1}\approx 2n\left(1-n\right)f_{2}$ and 
$m_{2}\approx nf_{2}$ (with $m_i\ll 1$).

\begin{figure}[bp]
  \centering
  \includegraphics[width=\columnwidth]{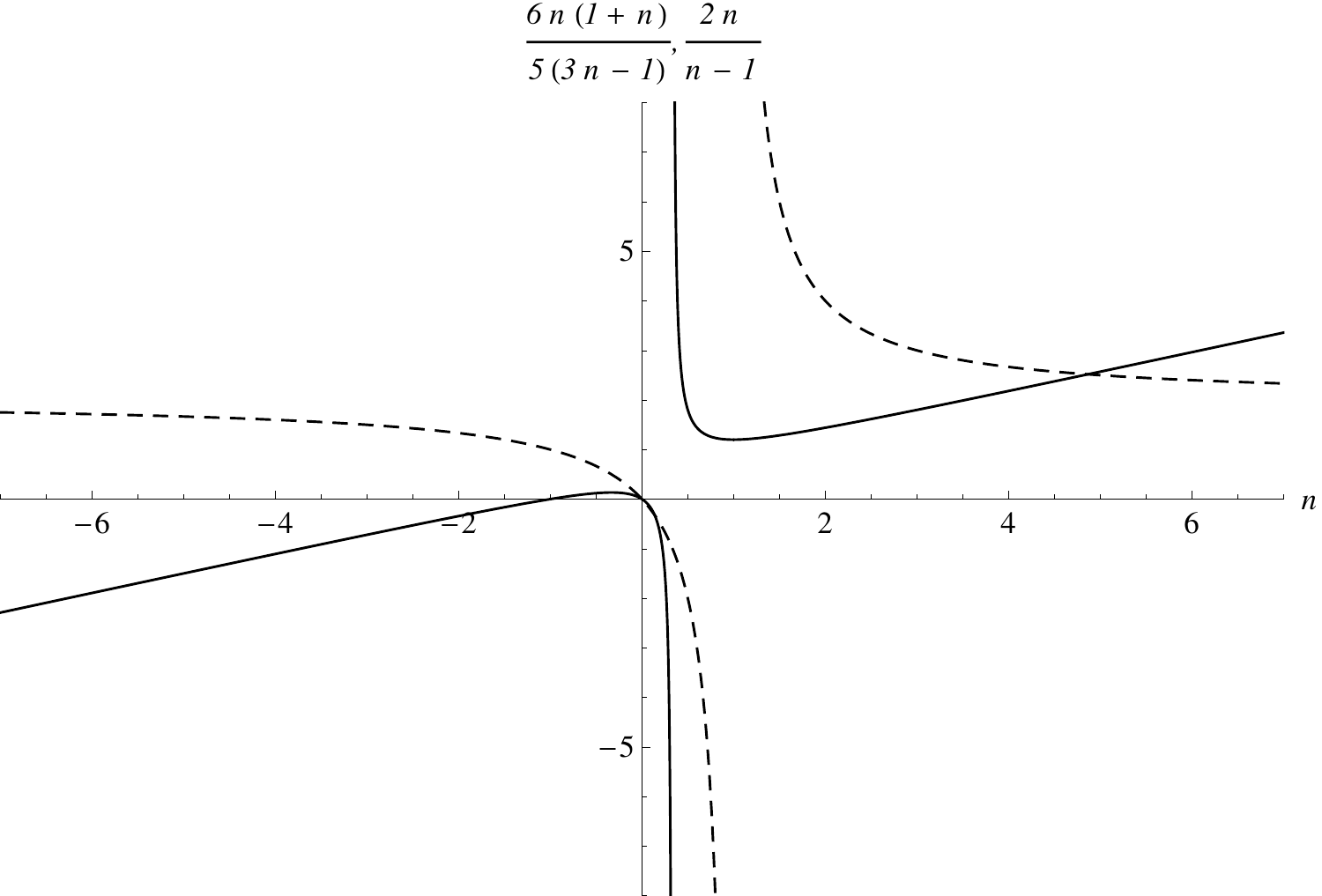}
  \caption{Growth behaviour induced by the presence of a non-minimal power-law coupling function with exponent $n$. The curves depict the effect of $n$ in Eqs. (\ref{delta_matter_era}) and (\ref{G_late_time}), for the matter dominated era with a perturbative (dark) or dominant coupling function (dashed).}
  \label{growth}
\end{figure}

At this stage the Universe is matter dominated, with $R\sim a^{-3}$ and $\Omega_m\approx 1$, and the equation for matter overdensities, rewritten here as a function of the number of \emph{e-folds} $N\equiv \ln a$, becomes

\begin{eqnarray}
{d^2\delta \over d^2N}+{1 \over 2}{d\delta \over dN}-{3\over 2}\left[1+A_n(k,a)\right]\;\delta_{m}&=&0~~,
\label{delta_N}
\end{eqnarray}

\noindent where

\begin{eqnarray}
A_n(k,a)&\equiv& {k^{2}\over a^{2}R}\left(m_{1}-4m_{2}\right)= -2n\left(1+n\right){k^{2}\over a^{2}R}f_{2}\sim a^{1-3n}~~,
\end{eqnarray}

\noindent and one has neglected 

\begin{equation}\omega_{eff}=2nf_2\ll {k^2\over a^2R}m_i\ll 1~~,\end{equation}

\noindent in the subhorizon approximation. Assuming that the density is a function of the growth factor, $\delta\sim \exp\left\{\int f_g\,dN\right\}$, and the	 approximation, $\left|{df_g\over dN}\right|\ll f^2_g$, one has the growing mode solution

\begin{equation}
\delta \sim a^{1+{6n\left(1+n\right)\over 5\left(3n-1\right)}{k^{2}\over a^{2}R\ln a} f_{2}}  ~~,
\label{delta_matter_era}
\end{equation}

\noindent which leads to the standard result, $\delta \sim a$, for a matter dominated Universe, in the absence of the coupling function, $f_2=0$. Therefore, the sign of the factor, ${6n\left(1+n\right)/ 5\left(3n-1\right)}$, determines whether the modification of gravity, in this regime, attenuates or enhances the growth of matter perturbations. The result is presented in Fig. (\ref{growth}), where it is shown that the exponents in the range $-1<n<0$ and $n>1/3$ lead to an increased growth, and for $n<-1$ to a decrease. 

Naively, one could expect a critical behaviour in the region near $n\approx 1/3$. However, one should first notice that, for  $n=1/3$, the coefficient $A_n$ becomes constant and the resulting evolution of the fluctuation is 

\begin{eqnarray}
\delta &\sim& a^{1-{6\over 5}n\left(1+n\right){k^{2}\over a^{2}R}f_{2}}~~,
\label{delta_ap}
\end{eqnarray}

\noindent which does not present such divergence issue.

\subsubsection{Curvature accelerated expansion}
\label{curvature_acceleration}

At the onset of the matter dominated era one may have instead, $m_i k^2/\gg a^2 R$, when the coupling function begins to dominate and the parameters $m_i$ become of order unity. In this case, the total coupling constant is given by 
 
\begin{eqnarray}
{G_{eff}\over G_{}}&\approx&{G_{c}\over G_{}}\approx-4{k^{2}\over a^{2}R}{m_{2}^{2}\over m_{1}}\Sigma\approx{2n\over n-1}{k^{2}\over a^{2}}{1\over \kappa\rho_{m}}~~,
\label{G_late_time}
\end{eqnarray}

\noindent and, with the energy density of matter at the background level, $\rho_m\sim a^{-3}$, one concludes that the interaction linearly increases with the scale factor $G_{eff}\sim a$. 

The equation of growth, Eq. (\ref{delta_a}), becomes

\begin{eqnarray}
{d^2\delta \over dN^2}+{1\over 2}{d\delta \over dN}+B_n(k,a)\,\delta_{m}&=&0~~,
\end{eqnarray}

\noindent with
\begin{eqnarray}
B_n(k,a)&\equiv &-6{k^{2}\over a^{2}R} {m_{2}^{2}\over m_{1}}f_{2}={6n^{2}\over 1-n} {k^{2}\over a^{2}R}f_{2}\sim a^{1-3n} ~~,
\end{eqnarray}

\noindent and, using the perturbation written as a function of the growth factor, one obtains, $f^2_{g}\approx B_n(k,a)$, as long as $n<1$. The growing mode solution for the matter overdensity runs with the scale factor as

\begin{eqnarray}
\delta\sim a^{{2\over \left(1-3n\right)\ln a}\sqrt{{6n^{2}\over 1-n}{k^{2}\over a^{2}R}f_{2}}}~~,
\label{delta_ap2}
\end{eqnarray}

\noindent where one must have $n<1$, for consistency. In this case, one sees that there is also an overgrowth for $n<1/3$, as in the case of a perturbative coupling --- otherwise, for $n>1/3$, the density perturbation would decrease with cosmic expansion, which  corresponds to a decaying mode and does not describe the correct cosmological behaviour.
Besides the value $n=1/3$, there is an additional critical behaviour for the exponent $n=1$, {\it i.e.} a linear coupling function, $f_2\sim R$. For this value, the parameter $m_1$, given in Eq. (\ref{m1_dominant}), vanishes, and the gravitational interaction becomes

\begin{eqnarray}
{G_{eff}\over G_{}}&\approx&{G_{c}\over G_{}}\approx-12\left({k^{2}\over a^{2}R}m_{2}\right)^{2}\Sigma ~~,
\end{eqnarray}

\noindent with $G_{eff}>0$, as $\Sigma$ is negative (cf. Eq. (\ref{sigma2})). However, $\Sigma$, the quantity which characterizes the weak lensing, must be positive, leading one to conclude that $n=1$ does not correspond to a satisfactory cosmological behaviour.

In conclusion, it is clear that the condition for a negative exponent, $n<0$, is necessary for a correct cosmological growth of matter density perturbations in the matter dominated era. In fact, if the coupling term is perturbative, the growth of the density perturbations is less than the standard $\delta\sim a$ by a small amount. 
However, when the coupling takes order unity values, the perturbation grows with the scale factor faster than in GR. In the opposite scenario, the one of a positive exponent, the dominant coupling function would decrease during cosmic expansion along with matter density perturbations and one would not obtain the observed LSS. Therefore, if one considers the modified Friedmann Eq. (\ref{modified_friedmann}), such that

\begin{eqnarray}
\Omega_m+\Omega_c&=&1~~, 
\end{eqnarray}

\noindent where one defines the curvature density parameter

\begin{eqnarray}
\Omega_c&\equiv & {\rho_c\over 3(F_1-2F_2\rho_m)H^2}~~,
\end{eqnarray}

\noindent one would expect, for this non-minimal coupling model, a transition from a matter dominated Universe, $\Omega_m\approx 1$, into a curvature dominated one, $\Omega_c\approx 1$, thus signalling that the Universe enters a phase of accelerated expansion \cite{expansion}.

In what concerns the background evolution, it was shown that the observation constraints on the deceleration parameter are compatible with negative values for the power-law exponents such as $n=-4$ and $n=-10$ \cite{expansion}. Therefore, it is expected that the cosmic evolution of perturbations, which favor negative values for the late-time dominance of a power-law coupling term, would be possible within this model of background expansion. One highlights that there is no fine-tuning involved in the assumed non-minimal coupling driving the background accelerated expansion of the Universe, {\it i.e.} a wide range of exponents $n$ is compatible with cosmographic reconstitutions of the evolution of the Universe (as discussed in \cite{expansion} and references within, namely \cite{Gong}).

\section{Conclusions}

In this work we have obtained the linearized field equations, Eqs. (\ref{mfe_00}) and (\ref{mfe_ij}), along with the equation of motion for matter overdensities in the presence of the non-minimal coupling between matter and geometry, Eqs. (\ref{delta_complete}) and (\ref{delta_m}).

In what concerns the cosmic evolution of density perturbations, this modified theory of gravity alters not only the gravitational coupling constant, but also induces a backreaction effective perturbation due to the presence of the extra force along with an additional frictional term, Eq. (\ref{phi_c}), which is similar to the one due to the expansion of the Universe (see Eq. (\ref{frictional})). This last term reveals that the non-minimal coupling modification in the background acts only in an indirect form and accounts for the adiabatic expansion of the Universe without an interchange of energy between matter and curvature, in particular, in  the evolution of matter overdensities. 

However, as can be seen in the resulting expressions for the coupling constants $\tilde{G}$ and $G_c$, Eqs. (\ref{Gtilde}) and (\ref{Gc}) respectively, it is the quantity $\Sigma$, introduced in order to quantify the deviation in the weak lensing, which better captures the deviation from GR.

We have also studied the modification that the presence of the non-minimal coupling induces on the evolution of matter overdensities --- in particular, for a power-law function, $f_2(R)\sim R^n$, in the subhorizon approximation, $k^2\ll a^2H^2$, and for a matter dominated era in comparison with the standard behaviour of $\delta\sim a$, in Eqs. (\ref{delta_ap}) and (\ref{delta_ap2}). 

The analysis for the regimes of a perturbative or dominant coupling function, in particular the obtained analytical solutions for the evolution of matter overdensities, lead to the conclusion that the conditions for a consistent cosmological behaviour require negative values for the power-law exponent. On one side, for the dominant coupling function, the weak lensing condition, $\Sigma>0$, along with the evolution of matter overdensities, Eq. (\ref{delta_ap2}), impose the conditions, $n<0$ (and $n<1/3$). On the other side, the necessity of a frictional attenuation in the growth of density perturbations, $\dot{\Phi}_c>0$, in both regimes, Eq. (\ref{phi_c}), only occurs for negative values of the power-law exponent. 

As posited in Ref. \cite{expansion}, a negative exponent naturally drives a transition from a matter dominated Universe to a phase of accelerated expansion. Hence, this work clearly shows that such a mechanism is compatible with the cosmological perturbations and the ensuing formation of LSS.

It should be highlighted that the non-minimal coupling does not need to be composed of a single power-law form, but can be considered to be a more evolved function, expressed as a sum of various terms,

\begin{eqnarray} 
f_2(R)&=&\sum_{n = - \infty}^{\infty} \left({R \over R_n}\right)^n ~~. 
\end{eqnarray}

\noindent where $R_n$ is a characteristic curvature value. Thus, different scenarios and scales (from astrophysical to cosmological) should lead to a dominance of one term over the others: for instance, in the context of this work, the dominant term must have a characteristic curvature $R_i$ of a cosmological scale. Naturally, other particular phenomena and environments can be described by this function, if the typical values of the curvature are such that other terms of the above series become relevant and induce the dynamics in each context: in particular, galactic dark matter can be accounted for by a power-law with exponent $n=-1$ or $n=-1/3$, with the curvature scales $R_{-1}$ and $R_{-1/3}$ of the order of $1/R_{DM}^2$, with $R_{DM}$ of the order of the radius of the dark matter halo \cite{mimic}. By studying a plethora of phenomena occurring at distinct scales, and thus probing different dominating power-law terms of the above expansion, one is then able to approximate the full behaviour the non-minimal coupling function.

\appendix
\section{First order perturbation of the field equations}
                                                                                                               
In the following, one introduces the linearization of the modified field equations, Eqs. (\ref{field_equations}), used to obtain Eqs. (\ref{mfe_00}) and (\ref{mfe_00}).

If each quantity contributes to a small deviation from the background value, the modified field equations, at first order, read

\begin{align}
\left(1+f_{2}\right)&\,\delta T_{\;\nu}^{\mu}+T_{\;\nu}^{\mu}\,\delta f_{2}=\nn\\
&=F\,\delta R_{\nu}^{\mu}+R_{\nu}^{\mu}\,\delta F-\frac{1}{2}\delta_{\nu}^{\mu}F_{1}\,\delta R\nn\\
&-\left(g^{\mu\rho}\bar{\nabla}_{\rho}\bar{\nabla}_{\nu}-\delta_{\nu}^{\mu}\bar{\square}\right)\delta F-\left[\left(\delta g^{\mu\rho}\right)\bar{\nabla}_{\rho}\bar{\nabla}_{\nu}+g^{\mu\rho}\nabla_{\rho}\nabla_{\nu}-\delta_{\nu}^{\mu}\square\right]F~~,
\end{align}
where $\bar{\nabla}_\mu$ stands for the background form of the covariant derivative and $\nabla_\mu$ the first order perturbation, and the same is assumed for the d'Alembert operator  $\square$.

For the time-time component one has

\begin{align}
F\delta R_{0}^{0}+&R_{0}^{0}(\delta F)-\frac{1}{2}F_{1}\delta R=\nn\\
&=-F\left[-\frac{k^{2}}{a^{2}}\Phi+3\ddot{\Psi}+3H(\dot{\Phi}+2\dot{\Psi})+3(\dot{H}+H^{2})(2\Phi)\right]+3(\dot{H}+H^{2})(\delta F)\nn\\
&-\frac{1}{2}F_{1}\left[-6(\dot{H}+2H^{2})(2\Phi)-6H(\dot{\Phi}+4\dot{\Psi})+6\ddot{\Psi}+2\frac{k^{2}}{a^{2}}(\Phi-2\Psi)\right]~~,
\end{align}

\noindent and

\begin{align}
-&(g^{00}\bar{\nabla}_{0}\bar{\nabla}_{0}-\bar{\square})\delta F-\left[(\delta g^{00})\bar{\nabla}_{0}\bar{\nabla}_{0}+(\nabla_{0}\nabla_{0}+\square)\right]F=\nn\\
&\qquad=-\ddot{F}(2\Phi)+(2\Phi)(\ddot{F}+3H\dot{F})+(\dot{\Phi}+3\dot{\Psi})\dot{F}-(\ddot{\delta F}+3H\dot{\delta F})+\frac{1}{a^{2}}\nabla^2(\delta F)\nn\\
&\qquad=3H\dot{F}(2\Phi)+(\dot{\Phi}+3\dot{\Psi})\dot{F}-3H\dot{\delta F}+\frac{1}{a^{2}}\nabla^{2}(\delta F)\;\;.
\end{align}

In the case of the spatial-spatial component, and since the background value of the function, $F$, depends only on time, the expression is given by

\begin{align}
F\delta R_{j}^{i}&+R_{j}^{i}\delta F-\frac{1}{2}\delta_{j}^{i}F_{1}\delta R\nn\\
-&(g^{ik}\bar{\nabla}_{k}\bar{\nabla}_{j}-\delta_{j}^{i}\bar{\square})\delta F-(\delta g^{ik})\bar{\nabla}_{k}\bar{\nabla}_{j}F-(g^{ik}\nabla_{k}\nabla_{j}-\delta_{j}^{i}\square)F\nn\\
&=F\delta R_{j}^{i}+R_{j}^{i}\delta F-\frac{1}{2}\delta_{j}^{i}F_{1}\delta R+\delta_{j}^{i}(\bar{\square}\delta F+\square F)-g^{ik}\bar{\nabla}_{k}\bar{\nabla}_{j}(\delta F)~~,
\label{mfe_ij_2}
\end{align}

\noindent where the first term is given by

\begin{align}
F\delta R_{j}^{i}+&R_{j}^{i}\delta F-\frac{1}{2}\delta_{j}^{i}F_{1}\delta R=\nn\\
&=\delta_{j}^{i}F\left[-(\dot{H}+3H^{2})\left(2\Phi\right)-H(\dot{\Phi}+6\dot{\Psi})-\ddot{\Psi}+\frac{1}{a^{2}}\nabla^{2}\Psi-\frac{1}{a^{2}}\delta^{ik}\bar{\nabla}_{k}\bar{\nabla}_{j}\left(\Phi-\Psi\right)\right]\nn\\
&\qquad+\delta_{j}^{i}(\dot{H}+3H^{2})\left(\delta F\right)\nn\\
&\qquad+\frac{1}{2}\delta_{j}^{i}F_{1}\left[6(\dot{H}+2H^{2})\left(2\Phi\right)+6H(\dot{\Phi}+4\dot{\Psi})+6\ddot{\Psi}+\frac{2}{a^{2}}\nabla^{2}\left(\Phi-2\Psi\right)\right]~~,
\end{align}

\noindent and one has used the spatial part of the curvature tensor for the metric, Eq. (\ref{metric}), which is given by

\begin{eqnarray}
R_{ij}&=&a^{2}\delta_{ij}\left[(\dot{H}+3H^{2})-(2\dot{H}+6H^{2})\left(\Phi+\Psi\right)-H(\dot{\Phi}+6\dot{\Psi})-\ddot{\Psi}+\frac{1}{a^{2}}\nabla^{2}\Psi\right]\nn\\
&&\qquad-\bar{\nabla}_{i}\bar{\nabla}_{j}(\Phi-\Psi)\;\;.
\end{eqnarray}

\noindent These results lead to the linearized field Eqs. (\ref{mfe_00}) and (\ref{mfe_ij}).


\end{document}